\documentclass[conference]{IEEEtran}
\usepackage{cite}
\usepackage{xcolor,soul,framed} 

\colorlet{shadecolor}{yellow}
\usepackage[pdftex]{graphicx}
\graphicspath{{../pdf/}{../jpeg/}}
\DeclareGraphicsExtensions{.pdf,.jpeg,.png}

\usepackage[cmex10]{amsmath}
\usepackage{array}
\usepackage{mdwmath}
\usepackage{mdwtab}
\usepackage{eqparbox}
\usepackage{url}
\usepackage{listings}
\usepackage{booktabs}
\usepackage{subfigure}
\usepackage{graphicx}
\usepackage{subcaption}
\usepackage{subfig}
\usepackage{float}
\usepackage{xcolor}
\usepackage{multicol}

\def\BibTeX{{\rm B\kern-.05em{\sc i\kern-.025em b}\kern-.08em
    T\kern-.1667em\lower.7ex\hbox{E}\kern-.125emX}}
    
\begin{document}

\title{Model, Analyze, and Comprehend User Interactions within a Social Media Platform} 

\author{\IEEEauthorblockN{Md Kaykobad Reza\IEEEauthorrefmark{1},
S M Maksudul Alam\IEEEauthorrefmark{2}, 
Yiran Luo\IEEEauthorrefmark{3},
Youzhe Liu\IEEEauthorrefmark{4} and 
Md Siam\IEEEauthorrefmark{5}}
\IEEEauthorblockA{\IEEEauthorrefmark{1}\IEEEauthorrefmark{2}\IEEEauthorrefmark{3}\IEEEauthorrefmark{4}Department of Computer Science,
University of California, Riverside\\
\IEEEauthorrefmark{5}Institute of Information Technology, University of Dhaka\\
\IEEEauthorrefmark{1}mreza025@ucr.edu,
\IEEEauthorrefmark{2}salam031@ucr.edu,
\IEEEauthorrefmark{3}yluo147@ucr.edu,
\IEEEauthorrefmark{4}yliu908@ucr.edu,
\IEEEauthorrefmark{5}bsse1104@iit.du.ac.bd
}}

\maketitle

\begin{abstract}
In this study, we propose a novel graph-based approach to model, analyze and comprehend user interactions within a social media platform based on post-comment relationship. We construct a user interaction graph from social media data and analyze it to gain insights into community dynamics, user behavior, and content preferences. Our investigation reveals that while 56.05\% of the active users are strongly connected within the community, only 0.8\% of them significantly contribute to its dynamics. Moreover, we observe temporal variations in community activity, with certain periods experiencing heightened engagement. Additionally, our findings highlight a correlation between user activity and popularity showing that more active users are generally more popular. Alongside these, a preference for positive and informative content is also observed where 82.41\% users preferred positive and informative content. Overall, our study provides a comprehensive framework for understanding and managing online communities, leveraging graph-based techniques to gain valuable insights into user behavior and community dynamics.
\end{abstract}

\begin{IEEEkeywords}
User interaction modeling, Social media analysis, Post-comment relationship, Strongly Connected Component
\end{IEEEkeywords}

\section{Introduction}
\IEEEPARstart{S}{ocial} media platforms have become integral components of our modern life \cite{sma0}, offering spaces for users to engage, share content, and interact with one another. Understanding the dynamics of user interactions within these platforms is important for various applications, including understanding user behavior \cite{sma2, sma3}, community management \cite{sma4, sma1}, content recommendation \cite{sok1, sok2} etc. In this study, we present a novel graph based approach on modeling, analyzing, and comprehending user interactions within a social media platform, focusing specifically on the post-comment relationship.

Availability of huge amount of social media data provides researchers with vast opportunities to explore user behavior and community dynamics. Previous studies have focused on analysing social media data from different aspects. This includes popularity prediction \cite{cao2020popularity}, sentiment analysis \cite{sentiment, singh2020sentiment}, user preference prediction for business decision making \cite{arrigo2021social}, public opinion analysis \cite{publicopinion} etc. However, how people connect and interact with each other and what are the things they collectively prefer as a community is still under-studied because of incompleteness and complexity of the data. 

To address these challenges, we propose a graph-based approach to model and analyze user behavior and community dynamics. We construct a user interaction graph based on post-comment relationships and examine various graph properties to understand both user and community perspectives. Our study explores the following key questions:
\begin{itemize}
    \item What does the interaction within a social media community look like based on post-comment relationships?
    \item Are there sub-groups, and how strongly are members connected within these sub-groups? 
    \item Who are the most popular users, and are they connected?
    \item What types of posts, comments, and content are preferred and become popular in the community?
\end{itemize}

We investigate user interaction within the community to identify clusters and behavioral patterns. By analyzing user activity, posts, and comments, we gain insights into user preferences and community dynamics. Our study also highlights trends in content that resonate most with the community.

Key findings reveal that 55.44\% of the community members are active, with 56.05\% of them form the largest community. Smaller clusters exist but 37.06\% of users are not strongly connected. Although there are 11,875 active members, only 0.8\% significantly contribute to community dynamics. We also observe temporal variations in activity, with spikes during admission and internship periods. Moreover, our analysis shows a correlation between user activity and popularity, as well as a strong preference for positive and informative content.

In summary, our graph-based approach offers a comprehensive framework for modeling and analyzing user interactions. It provides valuable insights into community dynamics, user behavior, and content trends, paving the way for enhanced understanding and management of online communities.

\begin{figure*}[th]
    \centering
    \subfigure[]{\includegraphics[width=0.24\textwidth]{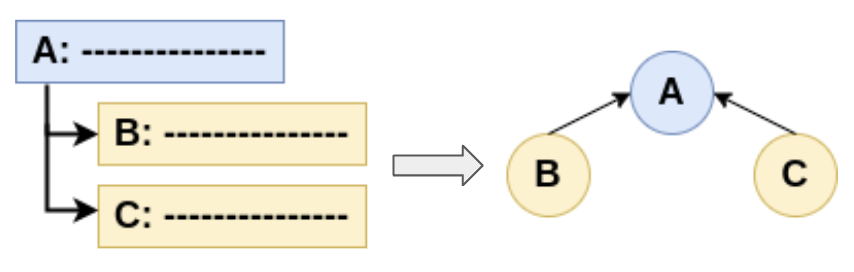}} 
    \hfill
    \subfigure[]{\includegraphics[width=0.24\textwidth]{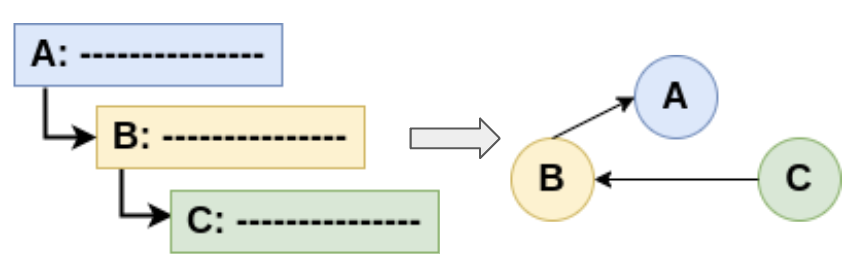}}
    \hfill
    \subfigure[]{\includegraphics[width=0.24\textwidth]{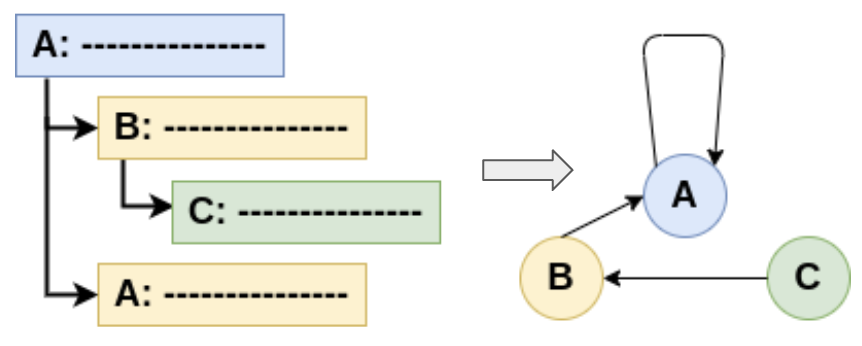}}
    \hfill
    \subfigure[]{\includegraphics[width=0.24\textwidth]{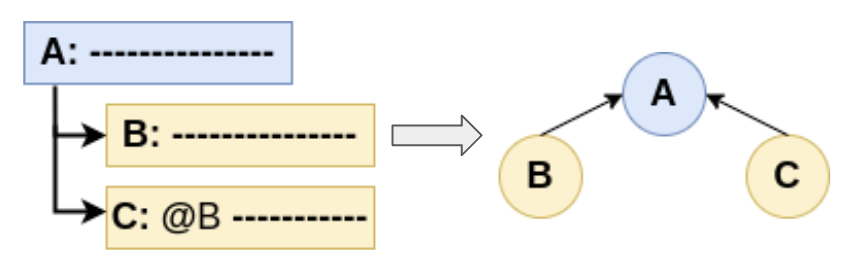}}
    \caption{(a) General commenting, ~~~~~~(b) Nested commenting, ~~~~~~(c) Self commenting, ~~~~~~(d) Tagged commenting.}
    \label{fig:graph-building}
\end{figure*}

\section{Related Work}
In social media analytics, particularly on reddit, numerous studies focus on understanding user engagement and content popularity. Kim et al. \cite{kim2021predicting} developed a machine learning model to predict reddit post popularity, finding neural networks most effective. Glenski et al. \cite{glenski2017predicting} recorded the behavior of 186 reddit users, presenting statistics on their interactions and finding simple models that could predict user behavior. Research also focuses on content, like the work from Barnes et al. \cite{barnes2021dank} analyzing 129,326 memes during COVID-19 \cite{covid, snigdha}, predicting meme popularity using machine learning.

Apart from popularity, studies on reddit data cover different aspects. Balsamo et al. \cite{balsamo2021patterns} investigate nonmedical opioid use, observing trends like synthetic opioids and rectal administration. Sawicki et al. \cite{sawicki2021exploring} highlight reddit as a data source for science, analyzing 180 manually annotated papers. Melton et al. \cite{melton2021public} conduct sentiment analysis on reddit discussions about COVID-19 vaccines, finding more positive than negative sentiments and focusing on side effects over conspiracy theories.

Reddit, a leading social media, is a popular source for data mining and analysis. Extensive research has been conducted on popularity \cite{cao2020popularity}, examining user behaviors \cite{sma0, sma3}, content moderation \cite{sok1, sok2}, and making business related decisions \cite{arrigo2021social}. These studies reveal patterns in user engagement and community formation. However, how users are intereconnected to each other and how they interact to form a community preference is still understudied. Our work builds upon this foundation, utilizing graph based approach to model complex user interaction, analyze different attributes and reveal useful insights to understand different community dynamics.

\section{Methodology}
This section outlines the data collection process, followed by the methodology for constructing and analyzing the user-interaction graph. Finally, it details the approach for analyzing user behavior and post-related insights.

\subsection{Data Collection}
We collected data from the r/ucr subreddit spanning January 2022 to December 2023 using the Arctic Shift tool\footnote{\url{https://arctic-shift.photon-reddit.com/download-tool}}, due to Reddit API limitations. The dataset includes comprehensive post attributes such as titles, contents, authors, comments, timestamps, and the post-comment hierarchy. This dataset is the foundation for our subsequent analyses of user interaction, behavior and community dynamics within the r/ucr subreddit. The dataset summary is presented in Table~\ref{tab:dataset-summary}. 

\begin{table}[th]
    \caption{Summary statistics of our dataset.}
    \centering
    \begin{tabular}{lr}
        \toprule
        \multicolumn{1}{c}{\textbf{Attribute}} & \multicolumn{1}{c}{\textbf{Value}} \\
        \midrule
        \midrule
        Number of Users in the Subreddit       & 21, 419                            \\
        Number of Active Users                 & 11,875                             \\
        Number of Posts                        & 18,037                             \\
        Number of Comments                     & 107,102   \\
        \bottomrule
    \end{tabular}
    \label{tab:dataset-summary}
\end{table}

\subsection{Building the User Interaction Graph} 
\label{BUIG}
This section outlines the construction of the user interaction graph based on post-comment relationships. We detail the methodology used to build the graph from the dataset and provide a critical analysis of its structural features and community clusters. Additionally, visualizations from various perspectives are presented to explore the network structure and key insights.

\subsubsection{Modeling Post-Comment Relationship with Graph}
We follow a number of principles, as illustrated in Figure~\ref{fig:graph-building}, while building the user-interaction graph. Each node represents a user and an edge indicates their connection through post-comment relationship. Specifically, for each comment made by a user, we draw a directed edge from the comment author to the post author, as shown in Figure~\ref{fig:graph-building}(a), ensuring appropriate flow of interaction. Furthermore, we connect each node to its immediate parent post/comment author rather than to the top-level post author as shown in Figure~\ref{fig:graph-building}(b). This ensures that interactions are traced back to their original context and hierarchy within the platform. Additionally, we introduce self-loops to nodes where authors comment on their own posts or comments as shown in Figure~\ref{fig:graph-building}(c). Finally, we do not consider explicit mentioning in comments; instead, we focus on the direct parent-child relationship as shown in Figure~\ref{fig:graph-building}(d). This ensures that the graph accurately reflects the immediate interactions between users. By adhering to these principles, our constructed user interaction graph provides a comprehensive representation of user engagement and interaction dynamics within the social media platform.

\subsubsection{Building the Network from Dataset}
\label{sec:building-uigraph}
The dataset is in CSV format with the following header:
\begin{lstlisting}
Author,author_fullname,created,downs,ups,
post_id,parent_id,permalink, Score,post,
title, subreddit_subscribers, 
upvote_ratio, post_name, Parent_post_author,
group_per_month,sentiment
\end{lstlisting}

This dataset contains detailed information about users and their posts and comments, enabling us to construct a network based on post-author and comment-author connections. We first create an adjacency list representation of the graph using the parent information of each post/comment. We then utilize the Python library NetworkX~\cite{hagberg2008exploring} to build the graph and Gravis\footnote{\url{https://robert-haas.github.io/gravis-docs/}} to visualize it.

\subsection{Clustering}
We analyze the User Interaction Graph (UIG) established in \ref{BUIG} to identify user clusters, or sub-groups, within the community. Our focus is on user-to-user connections formed through post-comment relationships. Each cluster is defined as a strongly connected component (SCC) within the UIG. We also measure close ties among users and redefine clusters based on these connections. To analyze the clusters, we create an adjacency list representation of the UIG and apply Tarjan’s algorithm \cite{tarjan} to identify SCCs. The results of the cluster analysis are presented in subsection~\ref{Res:CA}. 

\subsubsection{Weakly Formed Clusters (WC)}
\begin{figure}[t]
    \centering
    \subfigure[]{\includegraphics[width=0.14\textwidth]{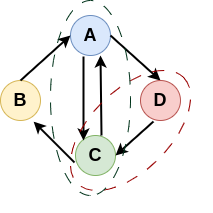}} 
    \subfigure[]{\includegraphics[width=0.14\textwidth]{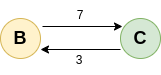}} 
    \subfigure[]{\includegraphics[width=0.14\textwidth]{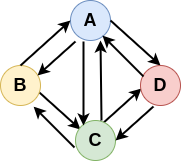}}
    \caption{(a) WC ~~~~(b) CTUP ~~~~(c) SC}
    \label{fig:clustering}
\end{figure}

In a Weakly Formed Cluster (WC), users in the UIG are connected to their neighbors through either direct or indirect connections. Our definition does not impose restrictions such as close ties between adjacent users or a minimum number of interactions when identifying clusters. Fig~\ref{fig:clustering}(a) illustrates an example of WC, where nodes 'A' and 'C' are strongly connected directly, while nodes 'C' and 'D' are connected indirectly through nodes 'B' and 'A'. Consequently, with this flexible definition, nodes 'A', 'B', 'C', and 'D' are considered members of a single cluster.

\subsubsection{Closely Tied User Pair (CTUP)}
We define Closely Tied User Pair (CTUP) based on interactions in the community, specifically through post-comment relationships. A pair of users is classified as a CTUP if one user has made three or more comments to the other. We set the threshold at three, assuming that repeated interactions are intentional. Fig~\ref{fig:clustering}(b) illustrates a CTUP, where user 'B' received three comments from user 'C', while user 'C' received seven comments from user 'B'. Additionally, we developed a formula to calculate the \textit{Tie-Score} between any two users in the community.
\begin{equation}
    \text{Tie-score} = \frac{\sum_{ij, ji}(ij, ji)}{0.4 \cdot \text{diff}(ij, ji)} 
    \label{eq:tie_score}
\end{equation}
 
Here, `ij' represents the number of comments from user `i' to user `j', while `ji' denotes the number of comments from user `j' to user `i'. We ranked all CTUPs in descending order of their \textit{Tie-Score} to identify the top closely tied user pairs.

\subsubsection{Strongly Formed Clusters (SC)}
Taking CTUPs into consideration, we imply a level of restriction to our definition of cluster. In Strongly Formed Clusters (SC), each of the adjacent user-pairs must have to be a CTUP ensuring that the adjacent users in a SC are closely tied by the post-comment relationship. Fig~\ref{fig:clustering}(c) shows an example where all the adjacent users such as: ('A','B'), ('A','D'), ('B','C') and ('C','D') are directly connected bidirectionally. The status of these pairs as CTUPs indicates a strong engagement level, reinforcing their close ties within the cluster and highlighting a cohesive community dynamic.

\begin{figure}[t]
  \centering
  \includegraphics[width=0.50\textwidth]{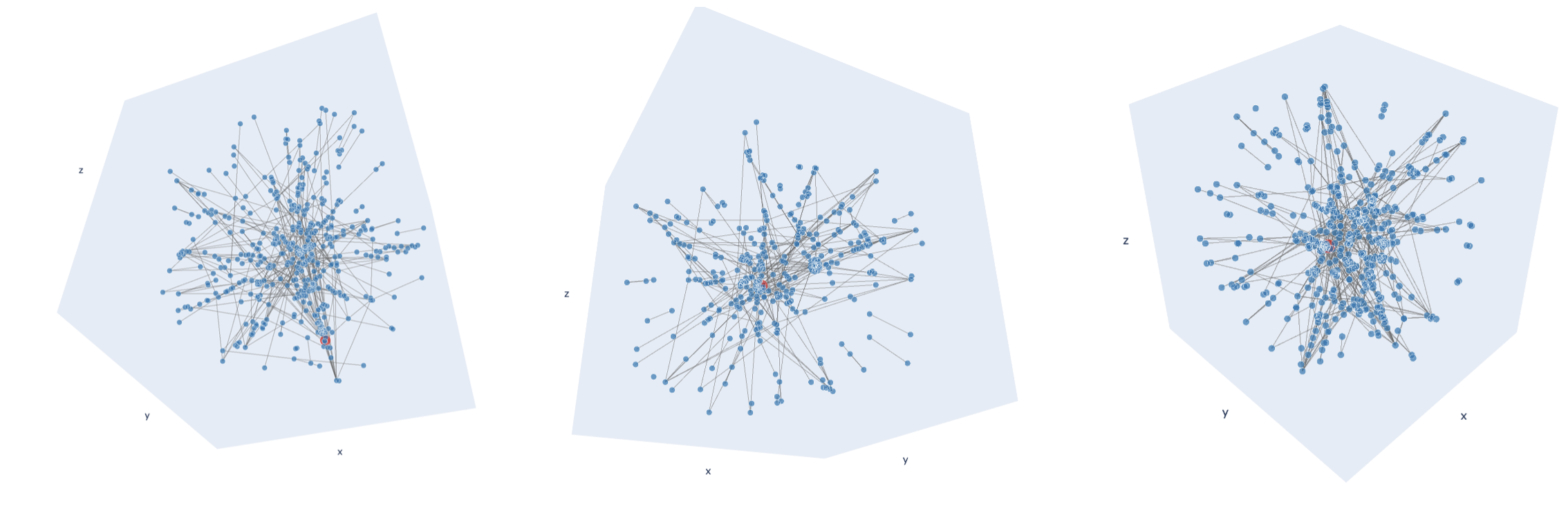}
  \caption{User interaction graph in different months}
  \label{fig:graph_months}
\end{figure}

\subsection{User and Post Analysis}
After collecting and pre-processing all the posts and comments from January 2022 to December 2023 from r/ucr subreddit, we analyze them in-depth from both user and post-comment perspectives. We process the data further and sort it according to different attributes. Then we go through the top results manually to understand the topic, content, context and user behavior properly. We use YAKE \cite{campos2020yake} library to automate the keyword extraction process. Our analysis has shown some interesting observations which we describe in Section~\ref{sec:results}.

\section{Result Analysis}
\label{sec:results}
In this section, we provide a detailed analysis of our findings. We begin with the analysis of the UIG, followed by a discussion of the clustering results. Finally, we highlight key insights from the analysis of both users and their posts.

\subsection{Analysis of the User Interaction Graph}
The UIG, constructed as described in Section~\ref{sec:building-uigraph}, eveals several intriguing insights. The forum is anonymous, so users don't know each other. They usually reply to posts that catch their interest. This leads to the formation of small communities around shared topics. For example, some users comment repeatedly on posts about similar topics.

To analyze further, we build graphs with different features to find out more details. We constructed graphs for different one-month periods, as shown in Figure~\ref{fig:graph_months}. We see that within a certain time period, the graphs are denser, and most of the users are connected. Since the r/ucr subreddit is relatively small, there are relatively few posts and comments each month. Additionally, trending topics often remain consistent over short timeframes, resulting in increased user interactions. In each period, we identify an influencer, highlighted in red, who possesses the highest degree.

\begin{figure}[t]
  \centering
  \includegraphics[width=0.47\textwidth]{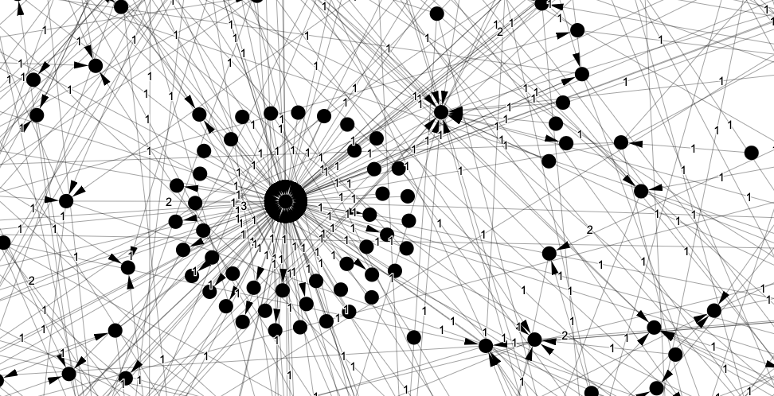}
  \caption{Network with edge weights}
  \label{fig:weighted graph}
\end{figure}

Next, we introduced weights to the graph to reflect the frequency of comments by users on each other's posts. A section of this weighted graph is illustrated in Figure~\ref{fig:weighted graph}. The graph is sampled from a subset of posts and comments for clearer visualization. 

Most edges have a weight of 1, indicating that users commented only once on the selected post or comment. However, some active users contribute multiple comments, leading to wdge weights of 2 or 3. This observation reinforces our earlier assumption that most users engage casually rather than fostering meaningful connections within the community. The high-density subgraphs correspond to trending posts and comments that attract increased user interactions. In light of these insights, we employ community detection algorithms to identify and visualize potential cohesive communities within the subreddit which we describe in the following section.

\subsubsection{Community Detection}
Community detection aims at identifying strongly connected groups or subgraphs within large-scale networks. We employed various algorithms, including centrality-based~\cite{freeman2002centrality}, Leiden~\cite{traag2019louvain}, and walktrap~\cite{pons2005computing} community detection.  Ultimately, we selected the centrality-based algorithm. It provides a straightforward and efficient approach to uncover community structures in the absence of clear divisions within the overall network.

\begin{figure}[t]
  \centering
  \includegraphics[width=0.46\textwidth]{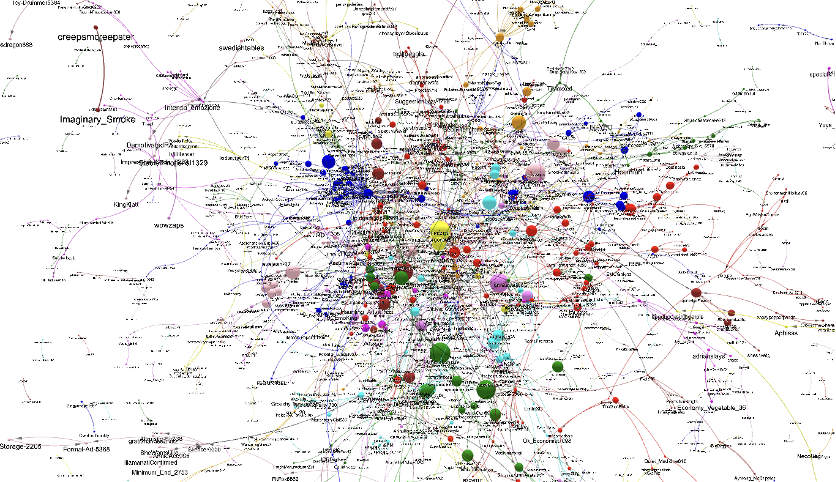}
  \caption{Community detection visualization}
  \label{fig:community}
\end{figure}

From Figure~\ref{fig:community}, we observe a large number of detected communities, indicating the absence of clear boundaries between substantial user groups. Most users participate in random activities rather than concentrating on specific friend groups, as seen on platforms like Twitter or Facebook. These sub-communities consist of users who frequently interact around specific topics or shared interests, reflecting a higher engagement level and the potential for meaningful discourse.

Our analysis also shows that, despite the overall trend of limited engagement, some users maintain one or two active connections within the network. This pattern suggests a nuanced interaction layer, where even less active users can significantly contribute to or benefit from specific threads.

\subsubsection{Summary of the Network}
Our analysis of the r/ucr subreddit reveals intriguing patterns of interaction on anonymous platforms. Most users engage minimally, contributing sporadic posts or comments. A few highly active users emerge as prominent figures, consistently participating in discussions. This behavior mirrors trends observed on platforms like Facebook and Twitter~\cite{onlinesocial, digitalsocial}. Additionally, we identified small groups where users engage in more frequent discussions, often centered around trending topics or popular posts, akin to gathering around an interesting event.

This study highlights that even in an anonymous forum, where users may seem transient, vibrant discussions occur. It resembles discovering small groups of friends conversing at a large party while most attendees merely pass by. Ultimately, our investigation reveals the various ways individuals can come together online, even if only briefly or around specific topics of shared interest. This highlights the subreddit as a dynamic space filled with diverse stories and connections.

\subsection{Cluster Analysis} \label{Res:CA}
Our analysis indicates that most users in the community are connected to others and are part of non-overlapping clusters.

\begin{itemize}
  \item 1 WC and 2 SCs: 7 users
  \item 2 SCs: 5 users
  \item 4 SCs: 4 users
  \item 9 WCs and 19 SCs: 3 users
  \item 51 WCs and 107 SCs: 2 users
\end{itemize}

We are able to identify 62 WCs and 135 SCs in-total within the UIG. The largest WC and SC contains 6,657 users and 624 users, respectively. Because of implying the CTUP restriction, the largest WC gets split into multiple smaller SCs. The data suggests that most users in the community are interconnected, either directly or indirectly. Additionally, it is evident that most clusters contain a small number of users.

\begin{figure*}[t]
    \centering
    \subfigure[]{\includegraphics[width=0.31\textwidth]{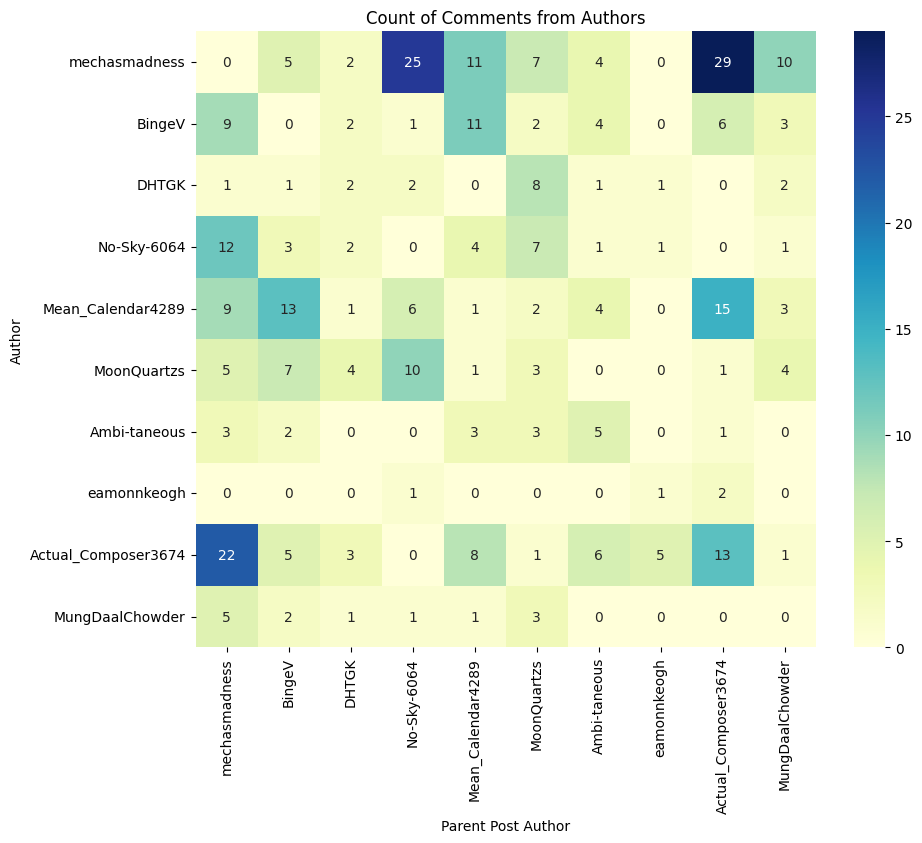}} 
    \hfill
    \subfigure[]{\includegraphics[width=0.31\textwidth]{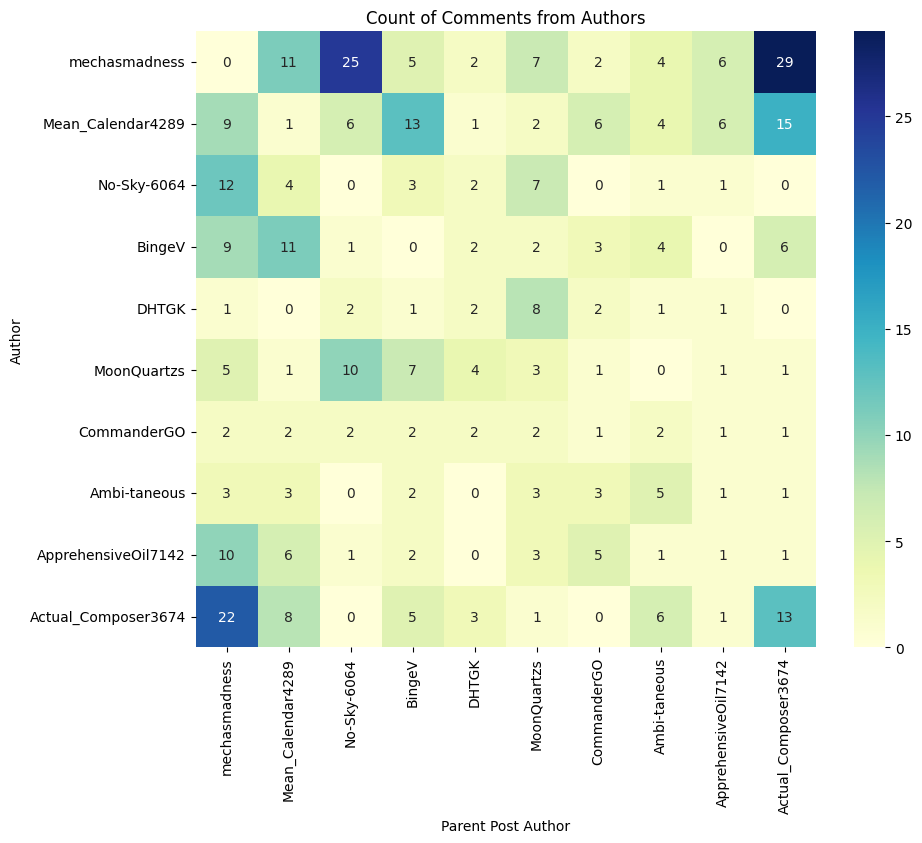}}
    \hfill
    \subfigure[]{\includegraphics[width=0.31\textwidth]{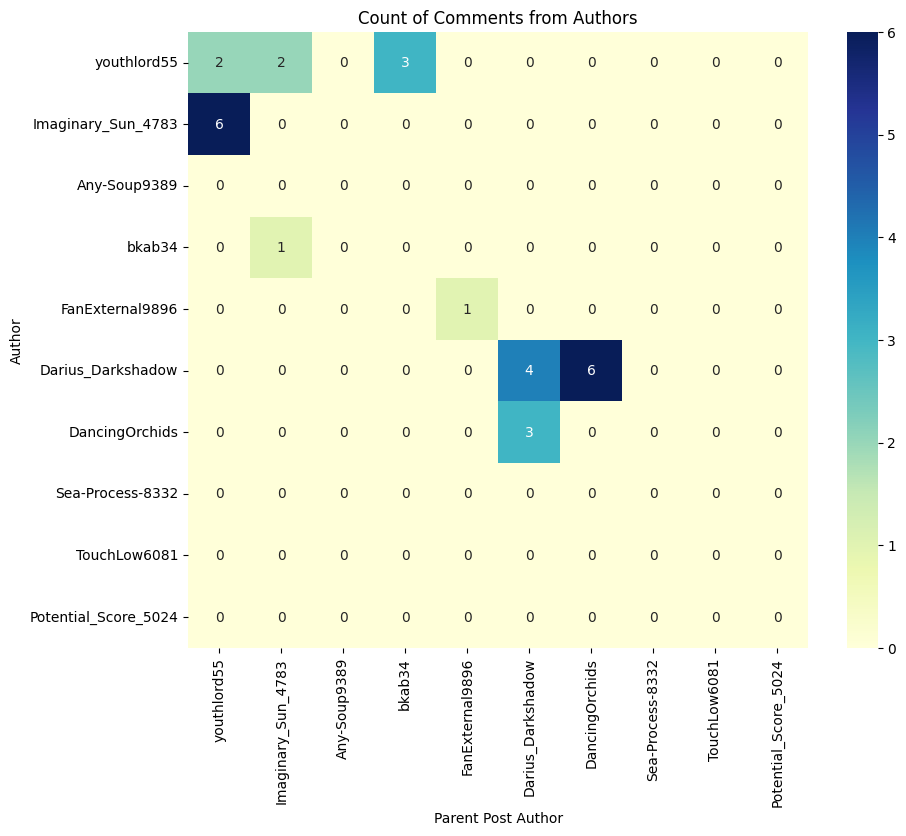}}
    \caption{Interaction between (a) Top 10 most upvoted users. (b) Top 10 most active users. (c) Top 10 most downvoted users.}
    \label{fig:interaction}
\end{figure*}

\subsection{User Analysis}
We conduct an in-depth analysis to identify the most influential and most disliked users in the community, along with the reasons behind their popularity and unpopularity. We aggregate the data to identify the top 10 most upvoted, top 10 most active and top 10 most downvoted users and manually analyzed their posts and comments. Our observations are summarized below:

\subsubsection{Top 10 Most Upvoted Users}
 Our findings indicate that users generally upvote and engage with content because they:
\begin{itemize}
    \item Share informative posts and comments
    \item Respond to questions with positive attitudes
    \item Present concise, accurate, and to-the-point content
    \item Occasionally incorporate humor
\end{itemize}

We also examine whether these users interact with one another. Figure~\ref{fig:interaction}(a) shows a heatmap revealing that the top 10 most upvoted users frequently comment on one another's content, indicating their connection through post-comment relationships within the same cluster.

\subsubsection{Top 10 Most Active Users}
Our analysis reveals a notable trend: more active users tend to be more popular, with increased activity correlating with a higher likelihood of receiving upvotes and comments. We observe approximately 80\% overlap between the top 10 most active users and the top 10 most upvoted users, suggesting that many active users are also highly upvoted. Similar to the top upvoted users, the most active users frequently comment on one another's posts, as shown in Figure~\ref{fig:interaction}(b).

\subsubsection{Top 10 Most Downvoted Users}
Our findings revealed that users generally downvote content from users who :
\begin{itemize}
    \item Create uninformative or irrelevant posts and comments
    \item Criticize others or display an arrogant attitude
    \item Engage in political discussions (disliked by students)
    \item Argue with others over trivial matters
\end{itemize}

Our investigation also shows that these users rarely comment on each other's content, indicating a lack of connection through post-comment relationships as shown in Figure~\ref{fig:interaction}(c).

\subsection{Post Analysis}
We conduct a thorough analysis of the posts and comments. We summarize our findings from four key perspectives:

\subsubsection{Top 10 Topics}
Our first focus is to identify the hot topics of discussion to gain valuable insights into the community's shared interests. Using the Python library YAKE\footnote{\url{https://liaad.github.io/yake/}}, we identified the top 10 topics in r/ucr subreddit which include:

\begin{multicols}{2}
\raggedright
1. Full-time student \\
2. Part-time student \\
3. Taking summer classes \\
4. Working part-time \\
5. School year starts \\
6. High school students \\
7. Classes you're taking \\
8. Working full-time \\
9. Financial aid office \\
10. Student account online
\end{multicols}

The first two topics pertain to full-time and part-time student statuses. Students often seek guidance regarding the transition between these statuses and ask questions on these topics. The subsequent topics can be categorized into four main areas: class taking, work and job, school affairs, and financial aid. These topics align with the primary concerns and interests of university students.

\subsubsection{Most Upvoted Posts}
We examine post titles, content, comments, and other pertinent details to identify the most upvoted posts. They can be categorized into three categories:
\begin{itemize}
    \item \textbf{Funny memes.} Posts featuring funny memes resonate well with the students, leading to high upvote counts.
    \item \textbf{Sharing of inspiring news.} Posts celebrating significant achievements, like graduation and admission, attract users to celebrate milestones within their academic community.
    \item \textbf{Safety concerns.} Posts addressing safety issues attract attention and prompt users to share experiences, suggestions, and concerns about campus safety.
\end{itemize}

\subsubsection{Most Commented Posts}
In our analysis of the most commented posts, we identified distinct categories that reveal unique insights. The categories are as follows:
\begin{itemize}
    \item \textbf{Sharing of inspiring news.} Similar to top upvoted posts, these posts resonate strongly within the community, as they reflect significant milestones for students.
    \item \textbf{Surveys and polls.} This category generates many comments due to users sharing opinions and connecting over mutual interests but receives fewer upvotes due to a lack of strong emotional responses.
    \item \textbf{Other Miscellaneous Topics.} Posts outside previous categories, like admission discussions or friendship-seeking, garner substantial commentary as users primarily engage to exchange information.
\end{itemize}

\subsubsection{Most Downvoted Posts}
We also examine the most downvoted posts to understand the commonalities that contributed to their negative reception. The posts are challenging to categorize due to the varied reasons for disapproval. Key themes identified include:

\begin{itemize}
    \item \textbf{Offensive speech.} Perceived as insensitive/inappropriate.
    \item \textbf{Selfish speech.} Disregard for community concerns.
    \item \textbf{Lack of empathy.} Disregard for others' feelings.
    \item \textbf{Patronizing tone.} Dismissive attitudes toward others.
    \item \textbf{Inappropriate humor.} Humor deemed unsuitable.
    \item \textbf{Oversimplification of facts and stereotypes.} Simplifying complex issues or relying on stereotypes.
\end{itemize}

These themes highlight the community's strong aversion to content that lacks sensitivity, empathy, and respect for diverse perspectives.

\section{Discussion, Challenges and Future Directions}
Our analysis shows that most users in the community prefer positive interactions. However, about 17.59\% of posts and comments include negative or aggressive language. These types of content were often downvoted by the community. Most discussions are focused on academic topics, jobs, and courses, which reflects the university environment. Overall, the atmosphere was mostly helpful and sometimes humorous.

We face several challenges while analyzing the data. The UIG is very dense and hard to read. We start by looking at the data month by month before switching to more straightforward methods. Though we use a few libraries to automate the analysis, but understanding user behavior and context still requires manual work. In the future, research could explore how influential users shape discussions and how algorithms recommendations affect user engagement. Additionally, studying how to promote positive interactions and reduce negativity would be helpful for community managers.

\section{Conclusion}
This study reveals important aspects of community dynamics. We find strong interconnectivity among users, with about 56.05\% of active members are part of a larger clusters. Notably, only around 0.8\% of users significantly influence community activity, indicating a power law distribution in engagement. Over two years, approximately 55.55\% of total members were active, and a correlation exists between user activity and popularity, showing that more active users tend to be more influential. The community favors positive content, with about 82.41\% of posts and comments being positive, which fosters a collaborative atmosphere. Overall, our findings offer insights into user behavior and content preferences that can guide community management strategies to enhance engagement and promote positive interactions.

\bibliographystyle{IEEEtran}
\bibliography{main}

\end{document}